# Observation of long phase-coherence length in epitaxial La-doped CdO thin films

Yu Yun[1,2,†], Yang Ma[1,2,†], Songsheng Tao[1,2], Wenyu Xing[1,2], Yangyang Chen[1,2], Tang Su[1,2], Wei Yuan[1,2], Jian Wei[1,2], Xi Lin[1,2], Qian Niu[1,3], X. C. Xie[1,2], and Wei Han[1,2*]

[1] International Center for Quantum Materials, School of Physics, Peking University, Beijing 100871, P. R. China
[2] Collaborative Innovation Center of Quantum Matter, Beijing 100871, P. R. China
[3] Department of Physics, University of Texas at Austin, Texas 78712, USA
[†]These authors contributed equally to the work
*Corresponding author: weihan@pku.edu.cn

**Abstract**

The search for long electron phase coherence length, which is the length that an electron can keep its quantum wave-like properties, has attracted considerable interest in the last several decades. Here, we report the long phase coherence length of ~ 3.7 micro meters in La-doped CdO thin films at 2 K. Systematical investigations of the La doping and the temperature dependences of the electron mobility and the electron phase coherence length reveal contrasting scattering mechanisms for these two physical properties. Furthermore, these results show that the oxygen vacancies could be the dominant scatters in CdO thin films that break the electron phase coherence, which would shed light on further investigation of phase coherence properties in oxide materials.

## I. INTRODUCTION

The electron phase-coherence length and time are the length and time that an electron can travel before losing its phase coherence, and are among the most important quantum, wave-like properties for one electron [1,2]. In low-dimensional mesoscopic systems, the interference of the electron phases leads to very interesting physical phenomena, including Aharonov–Bohm oscillations, universal conductance fluctuation, weak localization effects, etc. [3-6]. The search for long electron phase-coherence length, which is critical for quantum-coherent electronic devices, has attracted considerable interest in the last several decades. One of the central issues of fundamental importance is to understand the underlying mechanisms responsible for the loss of phase coherence [1,2,7]. For example, several important scattering mechanisms including electron-electron interaction, electron-phonon scattering, and magnetic or spin-orbit scattering have been taken into account to explain the loss of phase coherence [2,7,8]. To experimentally probe the phase-coherence properties, weak localization, a quantum increase of the electron resistivity, has been extensively used in various systems including metals, semiconductors, superconductors, and the emergent Dirac materials [2,7-15].

Recently, oxide electronics has attracted much scientific and technological attention since the discovery of the high-mobility two-dimensional electron gas (2DEG) in $SrTiO_3$ and its heterostructures [16-20]. Despite the high mobility, the electron phase-coherence length is only ~160 nm at 1.3 K, and ~280 nm at 12 mK [21,22], which limits the potential use of oxide materials for quantum-coherent electronic devices.

Here, we report the observation of long phase-coherence length in lanthanum-doped cadmium oxide thin films ($La_xCd_{1-x}O$) and the identification of the oxygen vacancies as the major phase-scattering sources in oxide thin films. A long phase coherence length ($L_\varphi$) of ~ 3.7 µm is observed in $La_{0.04}Cd_{0.96}O$ thin film at 2 K via the weak localization measurement, which is comparable to the value reported in ultra-high-mobility GaAs/GaAlAs 2DEG [10,23,24]. Contrasting scattering mechanisms for the electron mobility and the phase coherence are identified via the systematical studies of the La doping and temperature dependences. We show that the oxygen vacancies could be the dominant scatters in CdO thin films that break the electron phase coherence.

## II. EXPERIMENTAL DETAILS

The $La_xCd_{1-x}O$ thin films are epitaxially grown on (001) MgO substrates by co-evaporating La and Cd in diluted ozone using oxide molecular beam epitaxy (MBE-Komponenten GmbH; Octoplus 400) system with a base pressure lower than $1\times10^{-10}$ mbar. Prior to the film growth, the MgO substrates were annealed at 600 ºC for 1 h to clean the surface. Then ~5nm MgO buffer layer was grown by e-beam deposition with the substrate temperature holding at 250ºC and a growth rate of ~0.18 nm/min. This thin MgO buffer layer was used to achieve better crystalline quality $La_xCd_{1-x}O$ films. La and Cd elements were co-deposited in diluted ozone pressure of ~$1\times10^{-6}$ mbar from La and Cd thermal effusion cells with the substrate temperature at 200ºC. During growth, *in situ* reflective high energy electron diffraction (RHEED) is used to characterize the crystalline quality. To avoid degradation in air during the electron transport and phase-coherence measurements, an ~5 nm MgO thin film is deposited as the capping layer. The electron transport properties of the $La_xCd_{1-x}O$ thin films were measured via the van der Pauw method using the Physical Properties Measurement System (PPMS; Quantum Design).

## III. RESULTS AND DISCUSSION

## A. Structure and electron transport properties

Figures 1(a) and 1(b) show the RHEED patterns of the MgO substrate and the ~15nm $La_{0.040}Cd_{0.960}O$ epitaxial thin film. Sharp RHEED patterns (Fig. 1(b)) and x-ray diffraction results (Fig. 1(c)) indicate good crystalline properties of the $La_{0.040}Cd_{0.960}O$ thin film. The thickness of the film ~15 nm is determined from the fringe peaks around the CdO (002) main peak. The root-mean-square roughness is ~0.15 nm, obtained from an area of 2×2 $\mu m^2$ obtained using *ex situ* atomic force microscopy (Fig. 1(c) inset). As shown in Fig. 1(d), the electron mobility ($\mu$) of the $La_{0.040}Cd_{0.960}O$ thin film increases as temperature decreases, while it remains almost constant below 50K. This behavior could be well reproduced by a scattering-dependent theoretical model, which includes scatterings from ionized impurity and longitudinal optical (LO) phonon [25]. The carrier concentrations exhibit little variation as the temperature decreases from 300 to 2 K, as shown in Fig. 1(e).

## B. Long phase-coherence length in $La_{0.040}Cd_{0.960}O$

Weak localization measurements are carried out to investigate $L_\varphi$ as a function of the temperature in this $La_{0.040}Cd_{0.960}O$ thin film. Fig. 2(a) shows the representative magneto-conductance curves (open circles) at 2, 10, 30, and 50 K, respectively. The ratio of the resistance correction due to the weak localization over the total resistance ($\Delta R/R$) increases as the temperature decreases, as shown in Fig. 2(b) inset. This temperature dependence is in good agreement with the theoretical expectation of ln (T) dependence of the $\Delta R$ (red dashed line) for weak localization [1,5,26,27]. The Hikami-Larkin-Nagaoka theory is used to extract $L_\varphi$, which successfully describes the conductance correction due to quantum interference in two-dimensional systems [26]:

$$\Delta\sigma(B) = \frac{\alpha e^2}{2\pi^2 \hbar}\left[\psi\left(\frac{1}{2} + \frac{\hbar}{4eBL_\varphi^2}\right) - \ln\left(\frac{\hbar}{4eBL_\varphi^2}\right)\right], \tag{1}$$

where $\Delta\sigma(B)$ is the magneto-conductance that is due to weak localization, $e$ is the electron charge, α is the weak localization coefficient, $\hbar$ is the reduced Plank constant, $\psi$ is the digamma function, and $B$ is the magnetic field. The fitted curves (the solid lines in Fig. 2(a)) based on Eq. (1) are in good agreement with our experimental results. As shown in Fig. 2(b), $L_\varphi$ increases as the temperature decreases, and it reaches 3.7 µm at 2 K. Despite its extremely low electron mobility less than 350 cm$^2$/Vs, the electron phase coherence length in La$_{0.04}$Cd$_{0.96}$O thin film is comparable to that in the ultrahigh-mobility GaAs/GaAlAs 2DEG and graphene (Table Ⅰ) [10,13,14,23,24], which indicates that La$_{0.04}$Cd$_{0.96}$O thin film might be a promising candidate for future quantum-coherent electronic devices. Furthermore, the electron phase-coherence length is much longer than that in high-mobility SrTiO$_3$ and its heterostructures (Table Ⅰ) [21,22], presenting the fundamental question of the phase-scattering mechanisms.

Table Ⅰ. Comparison of $L_\varphi$ for La$_x$Cd$_{1-x}$O thin films with other semiconducting, two-dimensional and oxide materials.

| Materials | Phase-coherence length $L_\varphi$ (µm) |
|---|---|
| GaAs/GaAlAs 2DEG [10,23,24] | 3-24 |
| Graphene [13,14,28] | 0.5-5 |
| Black phosphorus [29,30] | 0.03-0.1 |
| Topological insulator [12,31] | ~0.5 |
| LaAlO$_3$/SrTiO$_3$ 2DEG [21] | ~0.16 |
| Doped SrTiO$_3$ [22] | ~0.28 |
| InO$_2$ [32,33] | 0.5 |
| ZnO [34] | ~0.15 |
| La$_{0.04}$Cd$_{0.96}$O (This work) | 3.7 |

**C. The role of La doping in the electron transport properties**

To determine the physical origin for this long phase-coherence length, we systematically vary the La doping in these films and examine the role of La doping for the electron phase coherence. $La_xCd_{1-x}O$ thin films of ~15 nm with various La doping ($x$) from 0 to 0.062 are grown using the oxide molecular-beam epitaxy. RHEED and x-ray diffraction measurements indicate good crystalline properties for all these thin films except x=0.062 [25]. Fig. 3(a) shows the temperature-dependence of the mobility for these films, and all temperature dependent behaviors can be well explained by ionized impurity and LO phonon scatterings [25]. Fig. 3(b) shows the carrier concentration as a function of the temperatures for all the films, which exhibits little variation from 300 to 2 K. This feature is similar to previous reports of La-doped oxide systems, such as $SrTiO_3$ and $BaSnO_3$ [18,35]. Fig. 3(c) shows the La doping dependence of the mobility at 2 K. The electron mobility initially increases as the La doping increases from 0 to 0.012, reaching the maximum value of ~527 $cm^2V^{-1}s^{-1}$, and then it decreases as the La doping further increases. In $La_xCd_{1-x}O$ thin films, La ($La^{3+}$) populates the Cd ($Cd^{2+}$) sublattices, which will generate an additional electron by replacing one Cd atom. Thus, higher carrier concentrations are observed for higher doping of La. These electron transport properties are consistent with previous reports on CdO thin films with Dy doping [36]. Fig. 3(d) shows the La doping dependence of the carrier concentration. As the La doping concentrations increase from 0 to 0.045, the carrier concentrations increase from $7.8\times10^{19}$ $cm^{-3}$ to $7.1\times10^{20}$ $cm^{-3}$. When the La doping concentration increases to 0.062, the carrier concentration decreases slightly to $6.5\times10^{20}$ $cm^{-3}$, which might be related to slightly poor crystal quality compared to other doping films [25]. The sheet resistance at 2 K exhibits a strong decrease as the La doping increases, as shown in the inset of Fig. 3(d).

**D. The role of La doping in the electron phase coherence properties**

The electron phase-coherence properties are investigated in these $La_xCd_{1-x}O$ thin films using the weak localization measurements. Fig. 4(a) shows the representative magneto-conductance curves at 2 K (open circles) for La doping with x=0, 0.008, 0.012, and 0.040, respectively. Clearly, a narrower dip is observed for a higher La doping, indicating a longer $L_\varphi$. To quantitatively determine $L_\varphi$ as a function of the La doping, we numerically fit the experimental results based on Eq. (1), indicated by solid lines in Fig. 4(a). The obtained $L_\varphi$ at 2 K is summarized in Fig. 4(b). $L_\varphi$ increases dramatically from ~0.1 to ~2 μm as the La doping increases from 0 to ~0.02. As the La doping further increases from ~0.02 to 0.04, $L_\varphi$ exhibits a modest enhancement. For much higher doping of La (x=0.062), shorter $L_\varphi$ is observed, which can be attributed to slightly poor crystal quality compared to other doping films [25].

These results present the critical role of La doping in the dramatic enhancement of $L_\varphi$ compared to the CdO films. One possible reason to account for this observation is that the La doping induces more carriers that can screen the phase-scattering sources, such as oxygen vacancies. To further study the role of La doping, we systematically measure $L_\varphi$ and the electron phase-coherence time ($\tau_\varphi$) as a function of temperature for these four $La_xCd_{1-x}O$ films with La doping of 0.040, 0.028, 0.003, and 0, respectively. Clearly, there is a major difference observed for higher and lower La doping films (Figs. 5(a) and 5(b)). For the $La_xCd_{1-x}O$ films with x=0 and 0.003, $L_\varphi$ increases as the temperature decreases from 50 to ~10 K, and then it saturates below ~ 10 K (Fig. 5(a)), whereas, for the $La_xCd_{1-x}O$ films with x=0.040 and 0.028, $L_\varphi$ increases monotonically as the temperature decreases from 50 to 2 K (Fig. 5(b)). $\tau_\varphi$ is estimated from the electron phase-coherence length and the diffusion constant ($D$) based on the relation $L_\varphi = \sqrt{D\tau_\varphi}$. Meanwhile, the diffusion constant can be obtained from the relation [37]:

$$D = \frac{1}{2}v_f \lambda = \frac{\pi \hbar^2}{m_e^* e^2} \frac{1}{R_S}, \quad (4)$$

where $v_f$ is the Fermi velocity, $\lambda$ is the mean-free path, $R_S$ is the sheet resistance, and $m_e^*$ is the effective electron mass, which is ~0.21$m_e$ ($m_e$: free electron mass) for the La-doped CdO thin films in our study [38]. Detailed information about the transport properties, the sheet resistance, mean-free path, 2D carrier density, and Fermi velocity can be found in the Supplemental Materials [25]. For the La$_x$Cd$_{1-x}$O films with x=0 and 0.003, $\tau_\varphi$ increases as the temperature decreases from 50 to ~10 K, and then it saturates below ~10 K (Fig. 5(c)). The saturation of $\tau_\varphi$ indicates the presence of strong phase-scattering centers. For the La$_x$Cd$_{1-x}$O films with x=0.028 and 0.040 (Fig. 5(d)), $\tau_\varphi$ increases monotonically as the temperature decreases from 20 to 2 K following the relation $\tau_\varphi \sim T^{-1}$ (red dashed line), which could be attributed to electron-electron scattering mechanism [1,2,39,40]. When the temperature is higher (20 - 50 K), the temperature dependence of $\tau_\varphi$ follows the relation $\tau_\varphi \sim T^{-2}$ (magenta dashed line), which could be attributed to electron-phonon scattering or large-energy-transfer electron-electron scattering mechanisms [2,32]. The crossover of $\tau_\varphi$ from the linear to quadratic temperature dependence is at ~ 20 K [25]. One major difference between the low-doping films and the high doping films is that carriers have different origins. Oxygen vacancies are the main source of carriers in relatively low-doping films, while most of the carriers in relatively high-doping films come from the La$^{3+}$ replacing Cd$^{2+}$. Oxygen vacancies could be the scattering centers that could break the electron phase coherence. For example, for nondoped CdO film (x=0), a tiny weak anti-localization feature is indeed observed below 10 K (Fig. 5(e)), while it is not shown for higher doping of La (Fig. 5(f)).

The weak anti-localization results for the nondoped CdO film indicate the presence of strong spin-orbit scattering centers that break the phase coherence. This observation agrees well with the saturation of the phase-coherence time and length below 10 K in Fig. 5(a) and 5(c). It also explains the slight increase of the magneto-conductance for higher temperature in Fig. 5(e). Furthermore, post-annealing in oxygen of the non-doped CdO film is performed to decrease the oxygen vacancies. An increase of the phase-coherence length and the mobility is observed, which further supports that the oxygen vacancies are the major phase-coherence scattering centers [25].

**E. Electron phase-coherence time vs momentum scattering time**

Our results show that the electron mobility and electron phase coherence exhibit totally different behaviors. First, the temperature dependence of these two physical properties are different. The electron mobility saturates below the temperature of ~ 50 K, whereas, the electron phase-coherence length monotonically increases as the temperature declines from 50 to 2 K. Besides, these two physical properties show contrasting La doping dependence. In the doping region of La from 0 to 0.012, the electron mobility increases from ~150 to ~550 cm$^2$/Vs, while electron phase coherence ($L_\varphi$ and $\tau_\varphi$) increases remarkably. In the doping region of La from 0.012 to 0.040, the electron mobility decreases from ~ 550 to ~ 350 cm$^2$/Vs. However, the electron phase coherence ($L_\varphi$ and $\tau_\varphi$) keeps increasing, which scales inversely with the electron diffusion constant. The La doping dependence of the momentum scattering time ($\tau_p$), estimated from the electron mobility ($\tau_p = \frac{m_e^*}{e}\mu$), and $\tau_\varphi$ are summarized in Figs. 6(a) and 6(b).

Clearly, these observations indicate the mechanisms that determine the electron mobility, one of electron's "particle-like" properties, and the phase coherence, one of electron's "quantum wavelike" properties, are different. To be specific, the electron mobility is determined by the elastic-scattering process, while $L_\varphi$ and $\tau_\varphi$ are related to the inelastic-scattering process. In the doping region of La from 0.011 to 0.040, the elastic scattering increases due to the La doping as impurities, which results in the decreasing of the electron mobility. However, the decrease of oxygen vacancies by La doping results in the large enhancement electron phase-coherence lengths and times. These results suggest that oxygen vacancies in La-doped CdO with low La doping can break the electron phase coherence and could be the dominant phase-coherence scattering sources in general oxide materials.

**IV. CONCLUSION**

In summary, we observe a long phase-coherence length of ~3.7μm in $La_{0.04}Cd_{0.96}O$ thin films at 2 K via the weak localization measurement, which is comparable to the value reported in ultrahigh-mobility GaAs/GaAlAs 2DEG [10,23,24]. Contrasting scattering mechanisms for the electron mobility and the phase coherence are identified via the systematical study of the La doping and temperature dependences. The oxygen vacancies are identified to be the dominant scattering sources in CdO thin films that break the electron phase coherence, which provides a unique insight for further investigation of phase-coherence properties in all oxide materials.

**ACKNOWLEDGMENTS**

We acknowledge the fruitful discussion with Haiwen Liu and Junren Shi. We also acknowledge the financial support from National Basic Research Programs of China (973

Program Grants No. 2015CB921104 and No. 2014CB920902) and National Natural Science Foundation of China (NSFC Grants No. 11574006 and No. 11704011).

**References:**

[1] B. L. Altshuler and A. G. Aronov, *Electron-Electron Interactions in Disordered Systems* North-Holland, Amsterdam,, 1985).
[2] J. J. Lin and J. P. Bird, Recent experimental studies of electron dephasing in metal and semiconductor mesoscopic structures. *J. Phys. Condens. Matter.* **14**, R501 (2002).
[3] S. Washburn and R. A. Webb, Aharonov-Bohm effect in normal metal quantum coherence and transport. *Adv. Phys.* **35**, 375 (1986).
[4] B. L. Altshuler, D. Khmel'nitzkii, A. I. Larkin, and P. A. Lee, Magnetoresistance and Hall effect in a disordered two-dimensional electron gas. *Phys. Rev. B* **22**, 5142 (1980).
[5] P. A. Lee and T. V. Ramakrishnan, Disordered electronic systems. *Rev. Mod. Phys.* **57**, 287 (1985).
[6] F. Pierre, A. B. Gougam, A. Anthore, H. Pothier, D. Esteve, and N. O. Birge, Dephasing of electrons in mesoscopic metal wires. *Phys. Rev. B* **68**, 085413 (2003).
[7] L. Saminadayar, P. Mohanty, R. A. Webb, P. Degiovanni, and C. Bäuerle, Electron coherence at low temperatures: The role of magnetic impurities. *Physica E: Low-dimensional Systems and Nanostructures* **40**, 12 (2007).
[8] F. Pierre and N. O. Birge, Dephasing by Extremely Dilute Magnetic Impurities Revealed by Aharonov-Bohm Oscillations. *Phys. Rev. Lett.* **89**, 206804 (2002).
[9] P. Mohanty, E. M. Q. Jariwala, and R. A. Webb, Intrinsic Decoherence in Mesoscopic Systems. *Phys. Rev. Lett.* **78**, 3366 (1997).
[10] J. A. Katine, M. J. Berry, R. M. Westervelt, and A. C. Gossard, Determination of the electronic phase coherence time in one-dimensional channels. *Phys. Rev. B* **57**, 1698 (1998).
[11] M. Liu, J. Zhang, C.-Z. Chang et al., Crossover between Weak Antilocalization and Weak Localization in a Magnetically Doped Topological Insulator. *Phys. Rev. Lett.* **108**, 036805 (2012).
[12] D. Kim, P. Syers, N. P. Butch, J. Paglione, and M. S. Fuhrer, Coherent topological transport on the surface of Bi2Se3. *Nat. Commun.* **4**, 2040 (2013).
[13] F. V. Tikhonenko, D. W. Horsell, R. V. Gorbachev, and A. K. Savchenko, Weak Localization in Graphene Flakes. *Phys. Rev. Lett.* **100**, 056802 (2008).
[14] S. Engels, B. Terrés, A. Epping, T. Khodkov, K. Watanabe, T. Taniguchi,B.Beschoten, and C. Stampfer, Limitations to Carrier Mobility and Phase-Coherent Transport in Bilayer Graphene. *Phys. Rev. Lett.* **113**, 126801 (2014).
[15] J. Liao, Y. Ou, H. Liu, K. He, X. Ma, Q.-K. Xue, and Y. Li, Enhanced electron dephasing in three-dimensional topological insulators. *Nat. Commun.* **8**, 16071 (2017).

[16] A. Ohtomo and H. Y. Hwang, A high-mobility electron gas at the LaAlO3/SrTiO3 heterointerface. *Nature* **427**, 423 (2004).

[17] J. Mannhart and D. G. Schlom, Oxide Interfaces—An Opportunity for Electronics. *Science* **327**, 1607 (2010).

[18] J. Son, P. Moetakef, B. Jalan, O. Bierwagen, N. J. Wright, R. Engel-Herbert, and S. Stemmer, Epitaxial SrTiO3 films with electron mobilities exceeding 30,000[thinsp]cm2[thinsp]V-1[thinsp]s-1. *Nat Mater* **9**, 482 (2010).

[19] H. Y. Hwang, Y. Iwasa, M. Kawasaki, B. Keimer, N. Nagaosa, and Y. Tokura, Emergent phenomena at oxide interfaces. *Nat. Mater.* **11**, 103 (2012).

[20] J. A. Sulpizio, S. Ilani, P. Irvin, and J. Levy, Nanoscale Phenomena in Oxide Heterostructures. *Annual Review of Materials Research* **44**, 117 (2014).

[21] D. Rakhmilevitch, M. Ben Shalom, M. Eshkol, A. Tsukernik, A. Palevski, and Y. Dagan, Phase coherent transport in SrTiO$_3$/LaAlO$_3$ interfaces. *Phys. Rev. B* **82**, 235119 (2010).

[22] S. W. Stanwyck, P. Gallagher, J. R. Williams, and D. Goldhaber-Gordon, Universal conductance fluctuations in electrolyte-gated SrTiO3 nanostructures. *Appl. Phys. Lett.* **103**, 213504 (2013).

[23] M. Ferrier, L. Angers, A. C. H. Rowe, S. Guéron, H. Bouchiat, C. Texier, G. Montambaux, and D. Mailly, Direct Measurementof the Phase-Coherence Length in GaAs/GaAlAs SquareNetwork. *Phys. Rev. Lett.* **93**, 246804 (2004).

[24] P. Roulleau, F. Portier, P. Roche, A. Cavanna, G. Faini, U. Gennser, and D. Mailly, Direct Measurement of the Coherence Length of Edge States in the Integer Quantum Hall Regime. *Phys. Rev. Lett.* **100**, 126802 (2008).

[25] See supplementary materials for detailed calculation of the tempereature-dependent mobility, structure characterization, temperature dependence of the sheet resistance, mean free path, and Fermi velocity, post-annealing in oxygen results, and the detailed analysis of the temperature dependent phase-coherent time.

[26] S. Hikami, A. I. Larkin, and Y. Nagaoka, Spin-Orbit Interaction and Magnetoresistance in the Two Dimensional Random System. *Progress of Theoretical Physics* **63**, 707 (1980).

[27] H.-Z. Lu and S.-Q. Shen, Finite-Temperature Conductivity and Magnetoconductivity of Topological Insulators. *Phys. Rev. Lett.* **112**, 146601 (2014).

[28] F. Miao, S. Wijeratne, Y. Zhang, U. C. Coskun, W. Bao, and C. N. Lau, Phase-Coherent Transport in Graphene Quantum Billiards. *Science* **317**, 1530 (2007).

[29] Y. Du, A. T. Neal, H. Zhou, and P. D. Ye, Weak localization in few-layer black phosphorus. *2D Materials* **3**, 024003 (2016).

[30] Y. Shi, N. Gillgren, T. Espiritu et al., Weak localization and electron–electron interactions in few layer black phosphorus devices. *2D Materials* **3**, 034003 (2016).

[31] H. Peng, K. Lai, D. Kong et al., Aharonov-Bohm interference in topological insulator nanoribbons. *Nat. Mater.* **9**, 225 (2010).

[32] C.-Y. Wu, B.-T. Lin, Y.-J. Zhang, Z.-Q. Li, and J.-J. Lin, Electron dephasing in homogeneous and inhomogeneous indium tin oxide thin films. *Phys. Rev. B* **85**, 104204 (2012).

[33] Y. J. Zhang, Z. Q. Li, and J. J. Lin, Electron-electron scattering in three-dimensional highly degenerate semiconductors. *Eur. Phys. Lett.* **103**, 47002 (2013).

[34] E. M. Likovich, K. J. Russell, E. W. Petersen, and V. Narayanamurti, Weak localization and mobility in ZnO nanostructures. *Phys. Rev. B* **80**, 245318 (2009).

[35] A. Prakash, P. Xu, A. Faghaninia, S. Shukla, J. W. A. III, C. S. Lo, and B. Jalan, Wide bandgap BaSnO3 films with room temperature conductivity exceeding 104 S cm-1. *Nat. Commun.* **8**, 15167 (2017).

[36] E. Sachet, C. T. Shelton, J. S. Harris et al., Dysprosium-doped cadmium oxide as a gateway material for mid-infrared plasmonics. *Nat. Mater.* **14**, 414 (2015).

[37] D. Natelson, *Nanostructures and nanotechnology* (Cambridge University Press, 2015).

[38] P. H. Jefferson, S. A. Hatfield, T. D. Veal, P. D. C. King, C. F. McConville, J. Zúñiga–Pérez, and V. Muñoz–Sanjosé, Bandgap and effective mass of epitaxial cadmium oxide. *Appl. Phys. Lett.* **92**, 022101 (2008).

[39] B. L. Altshuler, A. G. Aronov, and D. E. Khmelnitsky, Effects of electron-electron collisions with small energy transfers on quantum localisation. *J. Phys. C* **15**, 7367 (1982).

[40] H. Fukuyama and E. Abrahams, Inelastic scattering time in two-dimensional disordered metals. *Phys. Rev. B* **27**, 5976 (1983).

# Figure 1

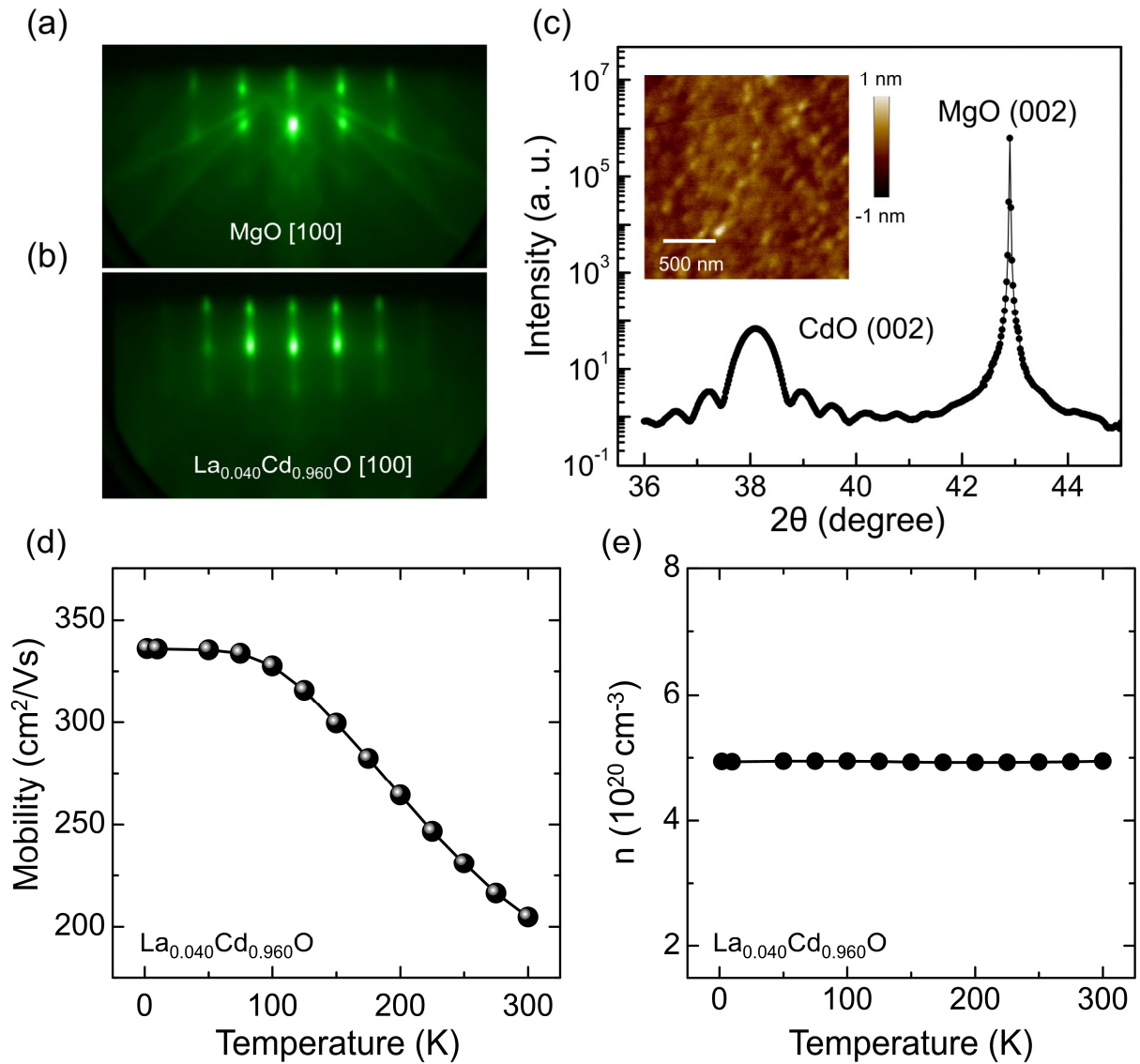

Figure 1. Growth and electron transport properties of the $La_{0.04}Cd_{0.96}O$ epitaxial thin film. (a)-(b) *In situ* RHEED characterization of the (001) MgO substrate and the $La_{0.04}Cd_{0.96}O$ epitaxial thin film (~15 nm) viewed along the [100] direction. (c) X-ray diffraction of the $La_{0.04}Cd_{0.96}O$ epitaxial thin film (~15 nm), where the thickness is determined from the fringe peaks around the CdO (002) main peak. Inset: Atomic force microscopy image of the $La_{0.04}Cd_{0.96}O$ film. (d)-(e) The temperature dependence of the electron mobility and carrier concentration of the $La_{0.04}Cd_{0.96}O$ film.

**Figure 2**

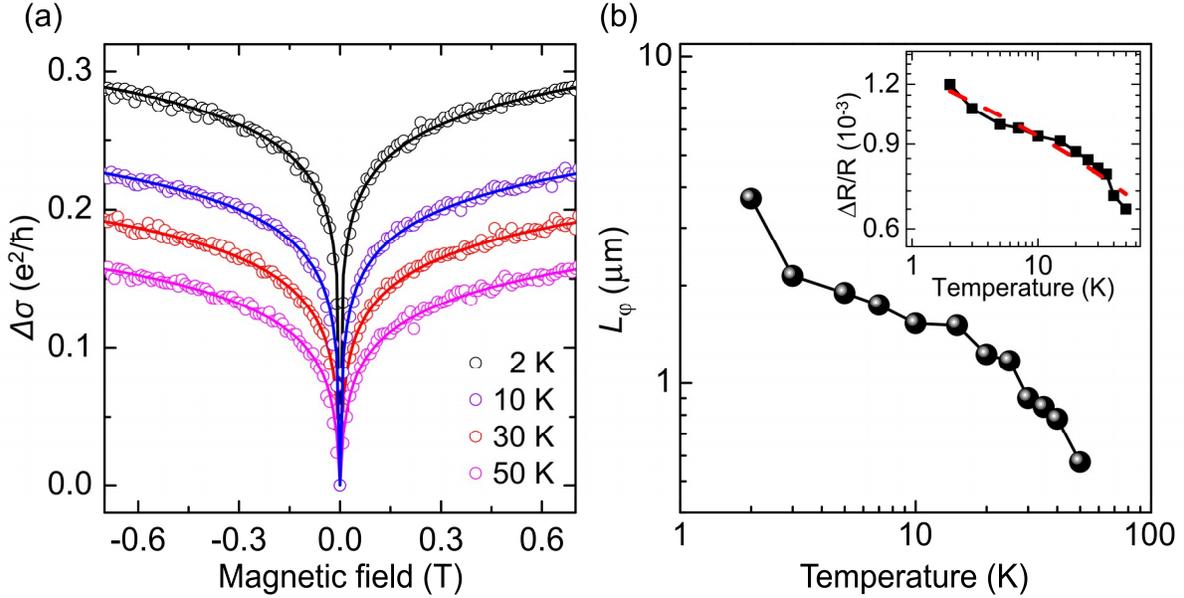

Figure 2. Temperature dependence of the electron phase-coherence length ($L\varphi$). (a) Representative weak localization results measured on the $La_{0.04}Cd_{0.96}O$ film in a perpendicular magnetic field at 2, 5, 10, 20, 30, and 50 K, respectively. The open circles indicate the experimental results, and the solid lines are the calculated results based on Eq. (1). (b) Temperature dependence of $L\varphi$ measured on the $La_{0.04}Cd_{0.96}O$ film. Inset: Ratio of resistance correction due to weak localization over total resistance ($\Delta R/R$) as a function of temperature. The red dashed line represents the ln ($T$) relationship.

**Figure 3**

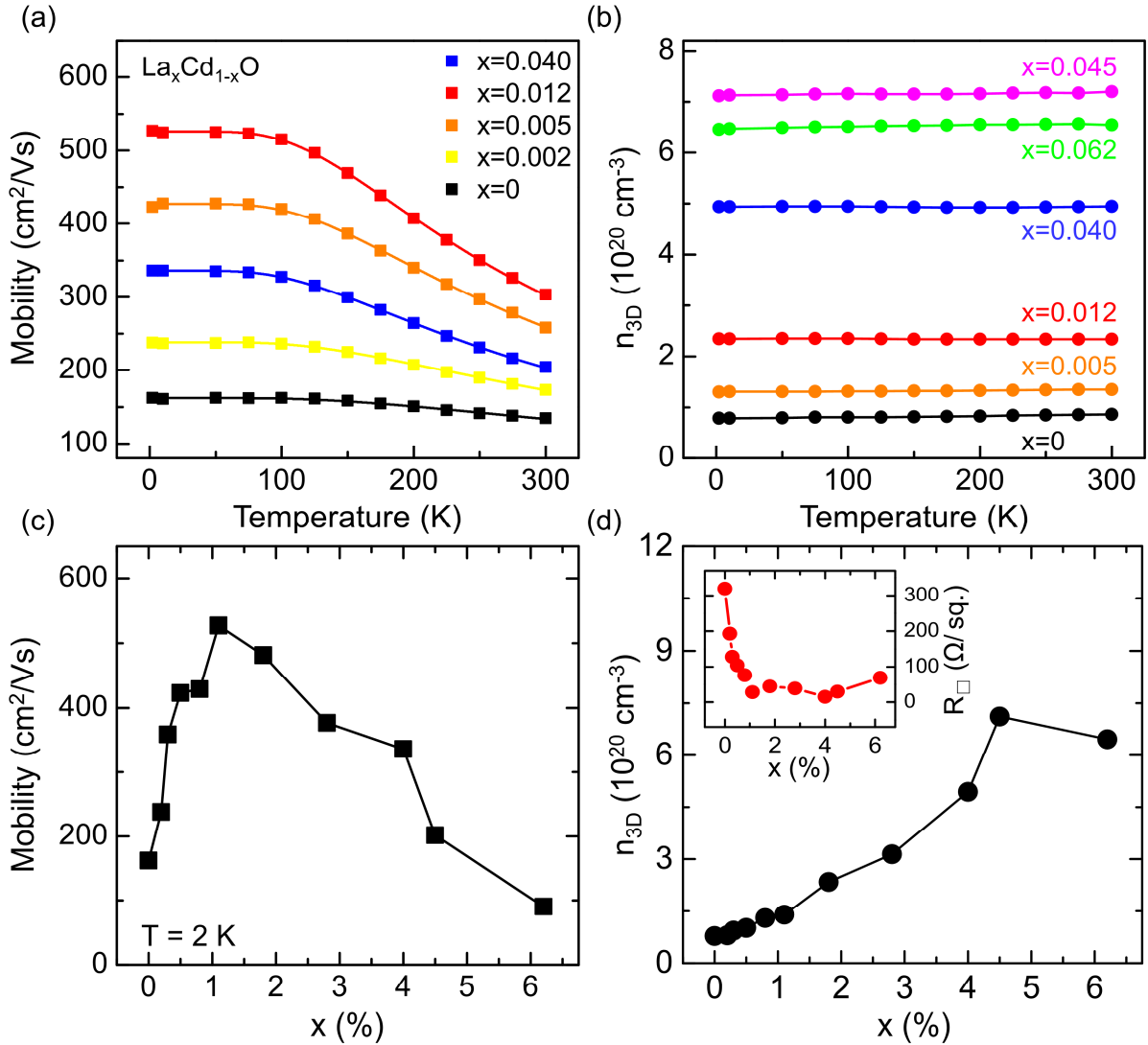

Figure 3. La doping dependence of electron transport properties for $La_xCd_{1-x}O$ thin films. (a)-(b) Temperature dependence of the mobility and carrier density for typical $La_xCd_{1-x}O$ thin films. (c)-(d) Mobility at 2 K and carrier density as a function of the La doping in $La_xCd_{1-x}O$ thin films. Inset of (d): Sheet resistance at 2 K as a function of the La doping.

**Figure 4**

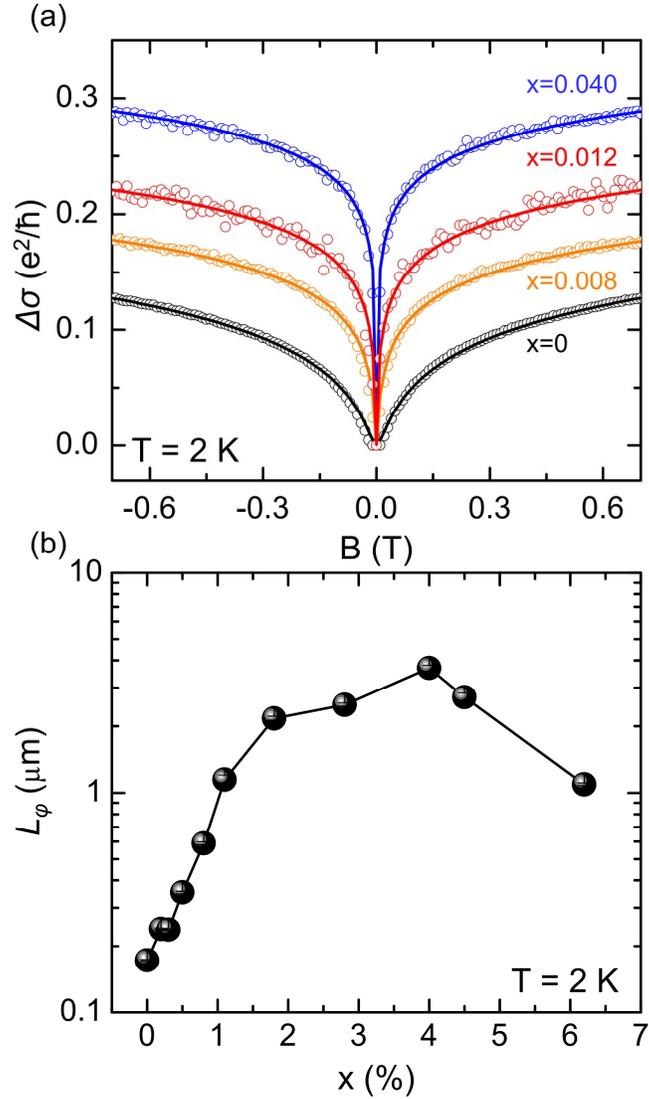

Figure 4. La doping dependence of phase-coherence properties for $La_xCd_{1-x}O$ thin films. (a) Representative weak localization results measured on the $La_xCd_{1-x}O$ thin films at 2 K with x=0, 0.008, 0.012, and 0.040, respectively. The open circles indicate the measured results, and the solid lines are the numerically calculated results based on Eq. (1). (b) La doping dependence of $L_\varphi$ at 2 K in $La_xCd_{1-x}O$ thin films.

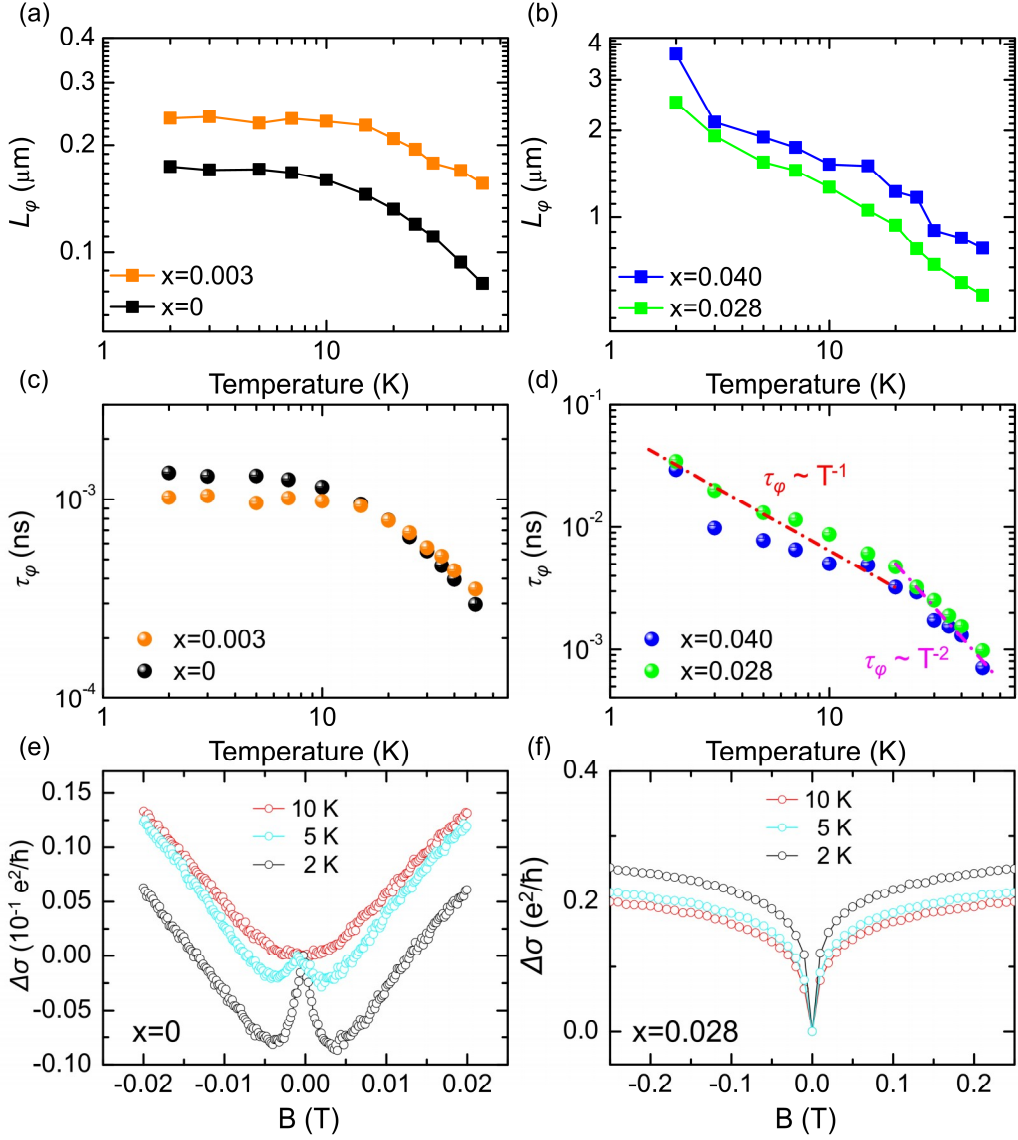

Figure 5. Temperature dependence of the phase-coherence lengths ($L_\varphi$) and the phase-coherence time ($\tau_\varphi$) for La$_x$Cd$_{1-x}$O thin films. (a)-(b) $L_\varphi$ vs temperature for La$_x$Cd$_{1-x}$O thin films with low doping of La (x=0 and x=0.002) and high doping of La (x=0.028 and x=0.040) respectively. (c)-(d) $\tau_\varphi$ vs temperature for La$_x$Cd$_{1-x}$O thin films with low doping of La (x=0 and x=0.002) and high doping of La (x=0.028 and x=0.040), respectively. The red and magenta dashed lines indicate the linear and quadratic temperature dependence. (e)-(f) Representative weak localization results measured on the La$_x$Cd$_{1-x}$O thin films at 2, 5 and 10 K with x=0 and x=0.028, respectively.

**Figure 6**

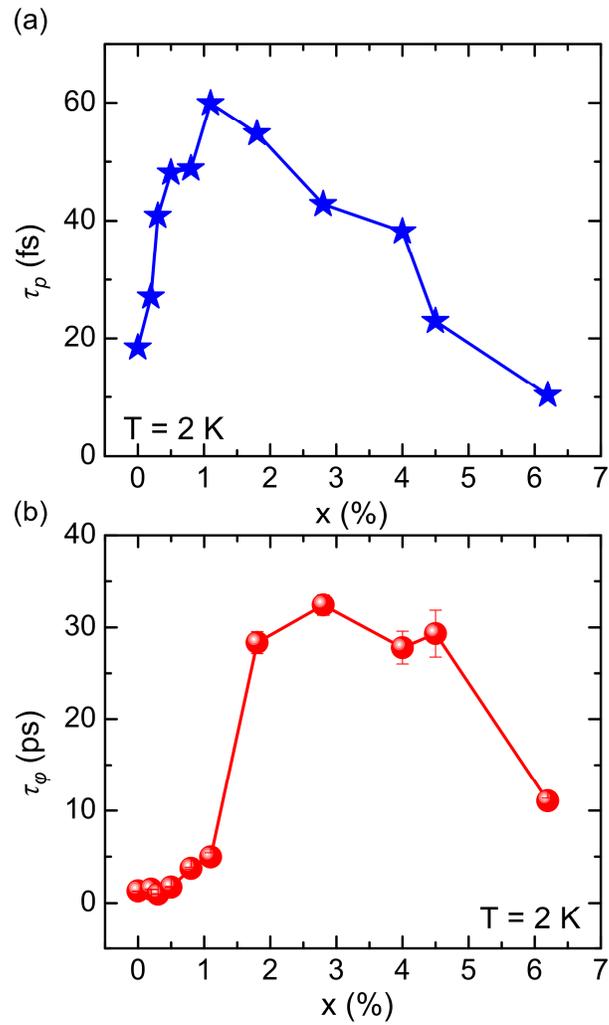

Figure 6. La doping dependence of the momentum scattering time (a) and electron phase-coherence time (b) of $La_xCd_{1-x}O$ thin films at 2 K.

# TABLE I

| Materials | Phase-coherence length $L_\varphi$ (μm) |
|---|---|
| GaAs/GaAlAs 2DEG [10,23,24] | 3-24 |
| Graphene [13,14,28] | 0.5-5 |
| Black phosphorus [29,30] | 0.03-0.1 |
| Topological insulator [12,31] | ~0.5 |
| LaAlO$_3$/SrTiO$_3$ 2DEG [21] | ~0.16 |
| Doped SrTiO$_3$ [22] | ~0.28 |
| InO$_2$ [32,33] | 0.5 |
| ZnO [34] | ~0.15 |
| La$_{0.04}$Cd$_{0.96}$O (This work) | 3.7 |

Table 1. Comparison of $L_\varphi$ for La$_x$Cd$_{1-x}$O thin films with other semiconducting, 2D, and oxide materials.

**Supplementary Materials for**

# Observation of Long Phase Coherence Length in Epitaxial La-doped CdO Thin Film

Yu Yun[1,2,†], Yang Ma[1,2,†], Songsheng Tao[1,2], Wenyu Xing[1,2], Yangyang Chen[1,2], Tang Su[1,2], Wei Yuan[1,2], Jian Wei[1,2], Xi Lin[1,2], Qian Niu[1,3], X. C. Xie[1,2], and Wei Han[1,2*]

[1] International Center for Quantum Materials, School of Physics, Peking University, Beijing 100871, P. R. China

[2] Collaborative Innovation Center of Quantum Matter, Beijing 100871, P. R. China

[3] Department of Physics, University of Texas at Austin, Texas 78712, USA

[†]These authors contributed equally to the work

*Correspondence to: weihan@pku.edu.cn

## S1. Temperature dependence of the electron mobility

The temperature dependent electron mobility can be attributed to ionized impurity and longitudinal optical (LO) phonon scattering mechanisms, as described in the equation below following previous work on SrTiO3 [1,2]:

$$\frac{1}{\mu(T)} = \frac{1}{\mu_{Impurity}} + \frac{1}{\mu_{Phonon}} \qquad (S1)$$

where $\mu_{Impurity}$ is the electron mobility due to ionized impurity, and $\mu_{Phonon}$ is the electron mobility due to LO phonon scattering, which has a strong temperature dependence, which could be described with a polaron model as previously derived by Low and Pines (42):

$$\mu_{Phonon} = \frac{1}{2\alpha w_l}\left(\frac{e}{m_p}\right)\left(\frac{m_e}{m_p}\right)^2 f(\alpha)\left(e^{\frac{\hbar w_l}{k_B T}} - 1\right) \qquad (S2)$$

where $\alpha$ is the electron-phonon coupling constant, $\hbar w_l$ is the LO phonon mode energy for scattering, $\hbar$ is the reduced Plank constant, $e$ is the electron charge, $m_p$ is the polaron mass, and $m_e$ is the electron effective mass, $f(\alpha)$ is a slowly varying function ranging from 1.0 to 1.4, $k_B$ is Boltzmann constant, and $T$ is the temperature.

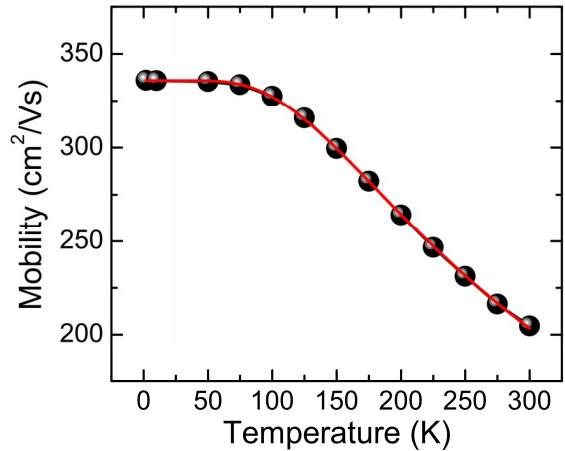

Fig. S1. Temperature dependence of the mobility for La$_{0.04}$Cd$_{0.96}$O. The red line indicates the fitting results of the mobility as a function of the temperature based on equations (S1) and (S2).

As shown in Fig. S1, the red solid line shows the fitting results based on equation (S1) and (S2), which is in a very good agreement with the experimental values (black circles) for the $La_{0.04}Cd_{0.96}O$ thin film. The obtained impurity mobility and phonon energy are 336 m²/Vs, and 38 meV, respectively. Furthermore, for all the $La_xCd_{1-x}O$ thin films with various La doping, the calculated mobility can well describe the measured mobility. Table S1 summarize the values for mobility due to ionized impurity and the LO phonon energies.

| $La_xCd_{1-x}O$ | Mobility at 2 K ($cm^2V^{-1}s^{-1}$) | LO phonon energy (meV) |
|---|---|---|
| 0 | 162 | 48 |
| 0.002 | 238 | 43 |
| 0.003 | 358 | 41 |
| 0.005 | 427 | 42 |
| 0.008 | 429 | 41 |
| 0.012 | 526 | 42 |
| 0.018 | 481 | 41 |
| 0.028 | 377 | 41 |
| 0.040 | 336 | 38 |
| 0.045 | 201 | 35 |
| 0.062 | 93 | 32 |

Table S1. The calculated values of the mobility due to ionized impurity and the LO phonon energies for $La_xCd_{1-x}O$ thin films.

## S2. Structure characterization

RHEED and XRD are used to characterize the structure properties of the $La_xCd_{1-x}O$ thin films, as shown in Fig. S2 and S3. For the high doping sample (x = 0.062), slightly poor crystal quality is observed compared to other doping films.

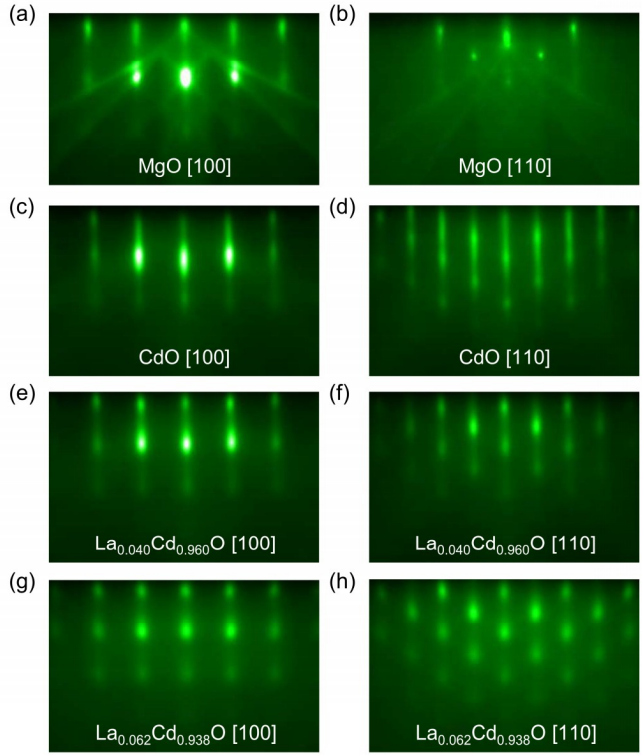

Fig. S2. RHEED patterns of the $La_xCd_{1-x}O$ epitaxial thin films. (a-b) RHEED patterns of the (001) MgO substrate viewed along the [100] and [110] directions, respectively. (c-h) RHEED patterns of the CdO, $La_{0.040}Cd_{0.960}O$, and $La_{0.062}Cd_{0.938}O$ epitaxial thin films (~ 15 nm) viewed along the [100] and [110] directions, respectively.

The (002) main peak of CdO shifts leftwards as La doping increases (Fig. S3(a)). The c-axis lattice parameter could be derived based on the XRD results (Fig. S3(b)). For the whole doping region, a tiny variation is observed ($\frac{4.74-4.715}{4.715} \times 100\% = 0.5\%$). Hence, the structural defect-induced elastic scattering can be ruled out to explain the variation of the phase coherence properties.

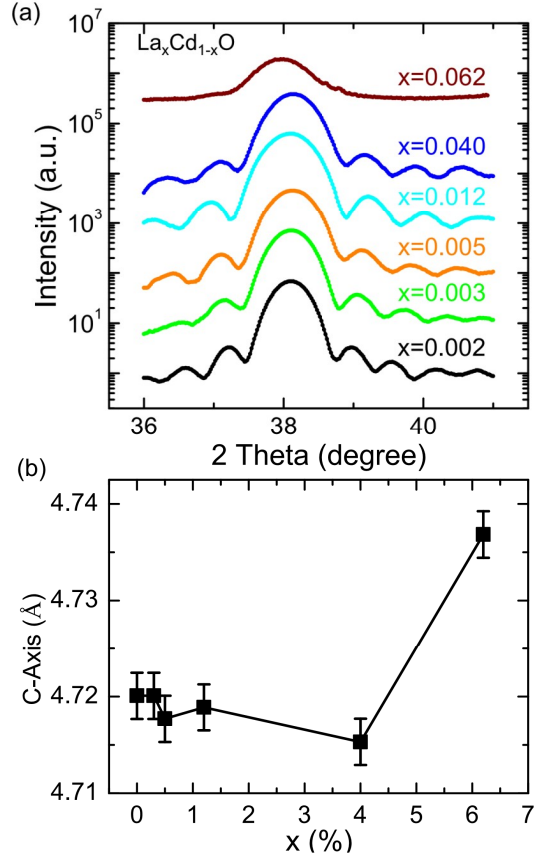

Fig. S3. X-ray diffraction characterization of the $La_xCd_{1-x}O$ epitaxial thin films. (a) Several representative X-ray diffraction curves of $La_xCd_{1-x}O$ on MgO substrates with x = 0.062, 0.040, 0.012, 0.005, 0.003, and 0.002, respectively. (b) The C-axis lattice constant as a function of the La doping.

## S3. Temperature dependence of the sheet resistance, mean free path, and Fermi velocity

$\tau_\varphi$ is estimated from the electron phase coherence length and the diffusion constant ($D$). During the calculation, detailed transport properties, the sheet resistance, mean free path, 2D carrier density, and Fermi velocity are shown in Fig. S4.

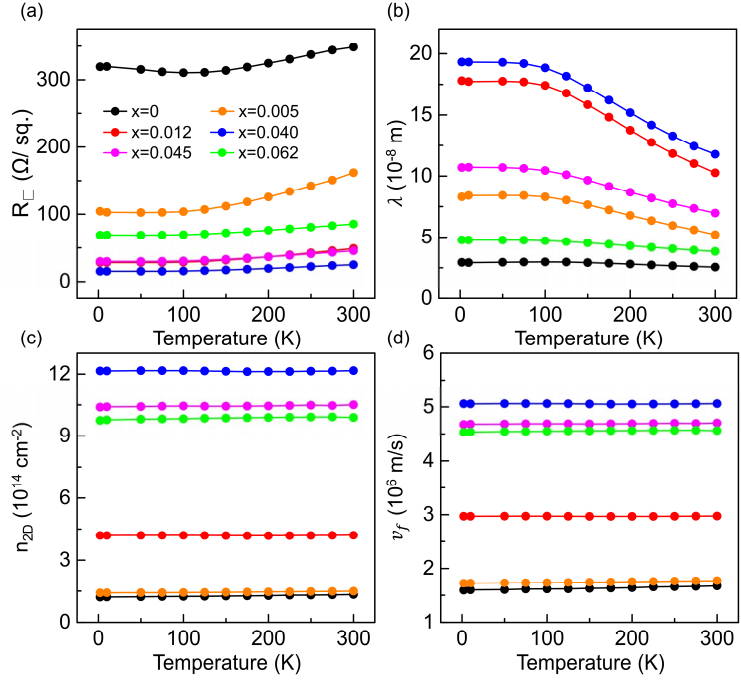

Fig. S4. Electron transport properties of $La_xCd_{1-x}O$ epitaxial thin films. (a-d) Sheet resistance($R_\square$), mean free path($\lambda$), sheet density($n_{2D}$) and Fermi velocity ($v_f$) for typical $La_xCd_{1-x}O$ films.

## S4. Post-annealing in oxygen gas

Regarding the role of the oxygen defects on the mobility and the phase coherence length, we have performed additional study on the non-doped films. By annealing the non-doped CdO film in oxygen gas at different temperatures, we observe an increase of both the electron mobility and phase coherence length, as shown in Fig. S5.

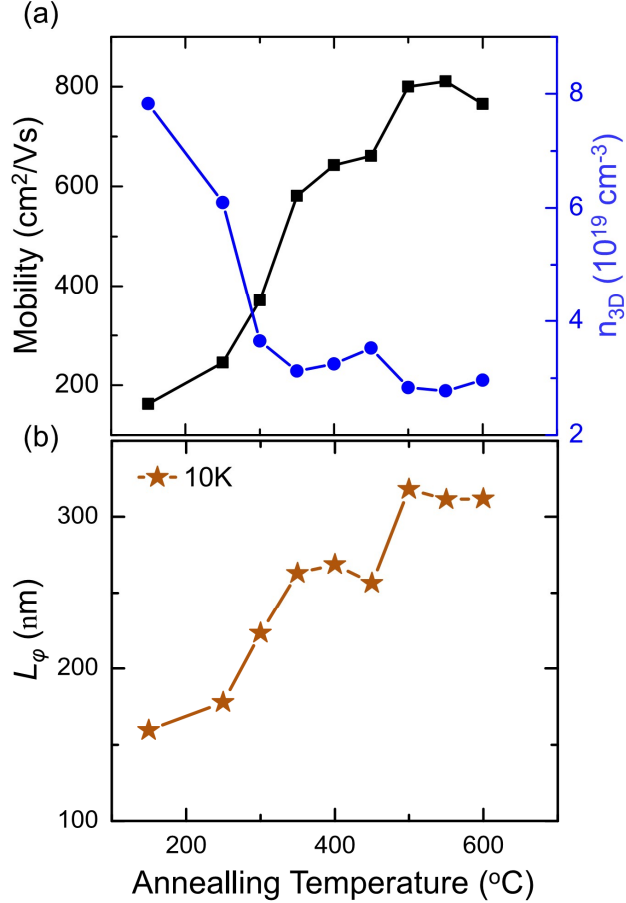

Fig. S5. The effect of post-annealing in oxygen gas on the electron transport and phase coherence length for the non-doped CdO epitaxial thin film. (a) Mobility and carrier density vs. the annealing temperature. (b) $L_\varphi$ vs. the annealing temperature.

## S5. Temperature dependence of the phase coherence time in Fig. 5 (d)

To determine the scattering mechanisms for the $La_xCd_{1-x}O$ films with x = 0.028 and 0.040, the Ln () vs. Ln (T) is plotted, as shown in Fig. S6. At low temperatures from 2 to 20 K, the slopes are -0.81 and -0.79 for x = 0.028 and x =0.040 samples. At high temperatures from 20 to 50 K, the slopes are -1.7 and -1.9 for x = 0.028 and x =0.040 samples. These results indicate a crossover from linear ($\tau_\varphi \sim T^{-1}$, electron-electron scattering) to quadratic temperature dependence ($\tau_\varphi \sim T^{-2}$, i.e. electron-phonon scattering or large-energy-transfer electron-electron scattering) for the dephasing rate. The crossover of $\tau_\varphi$ from the linear to quadratic temperature dependence is at ~ 20 K.

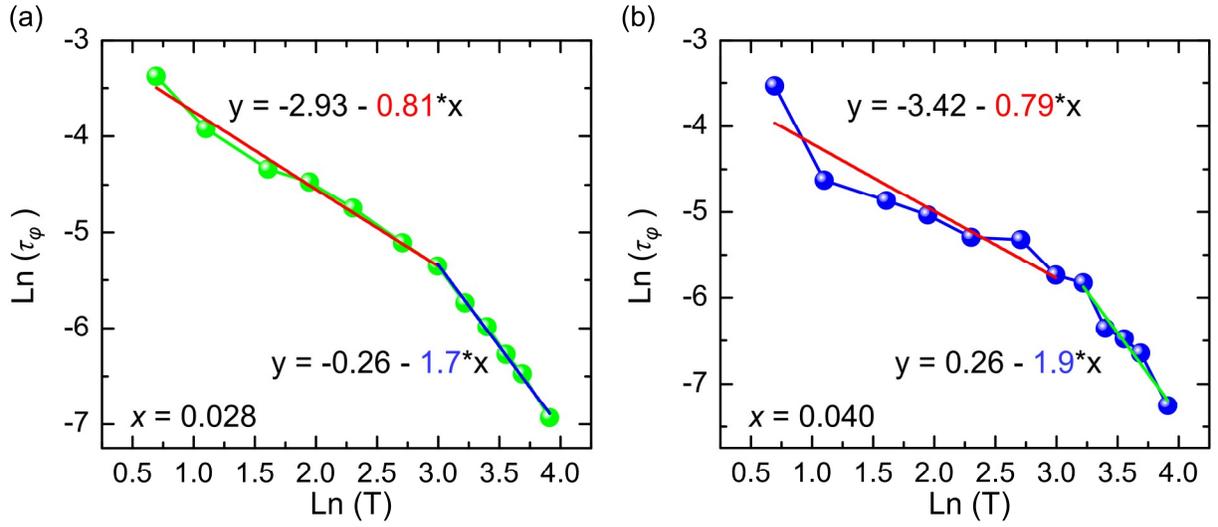

Fig. S6. The temperature of the phase coherence time for $La_xCd_{1-x}O$ thin films with doping of $x = 0.028$ (a) and $x = 0.040$ (b), respectively. The red, blue and green lines represent the best linearly fitting results. The phase coherence time and the temperature are in the units of ps and Kelvin.

**References:**

[1]    H. P. R. Frederikse and W. R. Hosler, Hall Mobility in SrTiO_{3}. *Phys. Rev.* **161**, 822 (1967).
[2]    A. Verma, A. P. Kajdos, T. A. Cain, S. Stemmer, and D. Jena, Intrinsic Mobility Limiting Mechanisms in Lanthanum-Doped Strontium Titanate. *Phys. Rev. Lett.* **112**, 216601 (2014).